\documentclass{article}

   \usepackage[nonatbib, final]{neurips_2023}

\usepackage[utf8]{inputenc} 
\usepackage[T1]{fontenc}    
\usepackage{hyperref}       
\usepackage{url}            
\usepackage{booktabs}       
\usepackage{amsfonts}       
\usepackage{nicefrac}       
\usepackage{microtype}      
\usepackage{lipsum}
\usepackage{graphicx}

\usepackage{amsmath}
\usepackage{esint}
\usepackage{subcaption}
\usepackage{comment}

\usepackage{booktabs}
\usepackage{amssymb}
\usepackage[version=4]{mhchem} 
\usepackage[colorinlistoftodos]{todonotes}
\usepackage{dcolumn}
\usepackage{tabularx}
\usepackage[export]{adjustbox}
\usepackage{makecell}
\usepackage{amsfonts}       
\usepackage{nicefrac}       
\usepackage{microtype}      
\usepackage{lipsum}
\usepackage{tipx}
\newcolumntype{C}[1]{>{\centering\let\newline\\\arraybackslash\hspace{0pt}}m{#1}}

\usepackage{nameref}
\usepackage{cleveref}
\usepackage{nomencl}

\usepackage{caption}

\newcommand{\angstrom}{\mbox{\normalfont\AA}}

\title{The Role of Reference Points in Machine-Learned Atomistic Simulation Models}

\author{
 Xiangyun Lei \\
  Energy and Materials Division\\
  Toyota Research Institute\\
  Los Altos, CA 94022 \\
  \texttt{ray.lei@tri.global} \\
   \And
 Weike Ye \\
  Energy and Materials Division\\
  Toyota Research Institute\\
  Los Altos, CA 94022 \\
  \texttt{weike.ye@tri.global} \\
   \And
 Joseph Montoya \\
  Energy and Materials Division\\
  Toyota Research Institute\\
  Los Altos, CA 94022 \\
  \texttt{joseph.montoya@tri.global} \\
  \And
 Tim Mueller\\
  Energy and Materials Division\\
  Toyota Research Institute\\
  Los Altos, CA 94022 \\
  \texttt{tim.mueller.ctr@tri.global} \\
  \And
 Linda Hung \\
  Energy and Materials Division\\
  Toyota Research Institute\\
  Los Altos, CA 94022 \\
  \texttt{linda.hung@tri.global} \\
  \And
 Jens Hummelshoej \\
  Energy and Materials Division\\
  Toyota Research Institute\\
  Los Altos, CA 94022 \\
  \texttt{jens.hummelshoej@tri.global} \\
}

\begin{document}

\maketitle

\begin{abstract}
This paper introduces the Chemical Environment Modeling Theory (CEMT), a novel, generalized framework designed to overcome the limitations inherent in traditional atom-centered Machine Learning Force Field (MLFF) models, widely used in atomistic simulations of chemical systems. CEMT demonstrated enhanced flexibility and adaptability by allowing reference points to exist anywhere within the modeled domain and thus, enabling the study of various model architectures. Utilizing Gaussian Multipole (GMP) featurization functions, several models with different reference point sets, including finite difference grid-centered and bond-centered models, were tested to analyze the variance in capabilities intrinsic to models built on distinct reference points. The results underscore the potential of non-atom-centered reference points in force training, revealing variations in prediction accuracy, inference speed and learning efficiency. Finally, a unique connection between CEMT and real-space orbital-free finite element Density Functional Theory (FE-DFT) is established, and the implications include the enhancement of data efficiency and robustness. It allows the leveraging of spatially-resolved energy densities and charge densities from FE-DFT calculations, as well as serving as a pivotal step towards integrating known quantum-mechanical laws into the architecture of ML models. 
\end{abstract}

\section{Introduction}
\label{sec:Introduction}
Atomic-scale simulations, including wavefunction-based methods and Density Functional Theory (DFT), have provided the predictive capabilities to enable computational drug discovery and materials design. However, due to computational cost and scaling, their application to large spatial and temporal scale problems can be a challenge. Machine learning, by contrast, has ushered in an era that promises to transcend these boundaries, potentially enabling the study of systems with millions or more atoms. However, machine learning force field (MLFF) models are predominantly grounded in atom-centered representations, hard-coding the resolution of the representation. As a result, the MLFF models similarly must also grapple with challenges like massive training data requirements and high computational demands. 

The Chemical Environment Modeling Theory (CEMT) introduces an approach that is more flexible than the atom-centered model, presenting a generalized framework that allows reference points (and features) to exist anywhere in the modeled domain. This framework not only addresses the current model constraints but also lays a foundation that may enable novel model architectures that are more performant or can target diverse applications.

\section{Background}
\label{sec:Background}
\begin{table}[h!]
  \centering
  \begin{tabular}{c|c|c}
    MLFF & CEMT & FE-DFT / FD-DFT \\
    \hline
    \begin{minipage}{0.3\textwidth}
        \begin{equation}
            E_{MLFF,sys} = \sum_{i=1}^{N_{atom}} e_{MLFF,i}
        \label{eq:MLFFPartition}
        \end{equation}
    \end{minipage} &
    \begin{minipage}{0.3\textwidth}
        \begin{equation}
            E_{CEMT,sys} = \sum_{i=1}^{N_{ref}} w_{i} \cdot e_{CEMT,i}
        \label{eq:CEMTPartition}
        \end{equation}
    \end{minipage} &
    \begin{minipage}{0.3\textwidth}
        \begin{equation}
            E_{DFT,sys} = \sum_{i=1}^{N_{grid}} w_{i} \cdot e_{DFT,i}
        \label{eq:DFTPartition}
        \end{equation}
    \end{minipage} \\
    \begin{minipage}{0.3\textwidth}
        \begin{equation}
            e_{MLFF,i} = f\circ \vec{\lambda}_{atom,i}(\vec{r},\vec{A})
        \label{eq:MLFFEnergyDensity}
        \end{equation}
    \end{minipage} &
    \begin{minipage}{0.3\textwidth}
        \begin{equation}
            e_{CEMT,i} = f \circ \vec{\lambda}_{ref,i}(\vec{r}, \vec{A})
        \label{eq:CEMTEnergyDensity}
        \end{equation}
    \end{minipage} &
    \begin{minipage}{0.3\textwidth}
        \begin{equation}
            e_{DFT,i} = f \circ  \vec{\lambda}_{grid,i} \circ \rho(\vec{r}, \vec{A})
        \label{eq:DFTEnergyDensity}
        \end{equation}
    \end{minipage} \\
     \\
  \end{tabular}
\label{tab:EquationComparison}
\caption{Basic equations for the theories of Machine Learning Force Fields (MLFF), Chemical Environment Modeling Theory (CEMT), and the electronic energy of real-space orbital-free Finite Element DFT (FE-DFT) or Finite Difference DFT (FD-DFT). $\circ$ denotes ``function composition'', e.g. $(g \circ f)(x)=g(f(x))$. $E_{sys}$ denotes system level total energy, $e_i$ denotes energy contribution, $w_i$ denotes weight, $\vec{\lambda}_{i}$ denotes feature set used to describe the local chemical environments, $f$ denotes function that predict energy contribution from features, $\vec{r}, \vec{A}$ are the relative positions and types of neighboring atoms, respectively, and $\rho$ denotes electron density distribution.  }
\end{table}

\begin{table}[h!]
  \centering
  \begin{tabular}{c|c}
    MLFF & CEMT \\
    \hline
    \begin{minipage}{0.45\textwidth}
        \centering
        \begin{equation}
        \begin{aligned}
             &F_{k,\alpha} = -\frac{\partial E_{MLFF,sys}}{\partial R_{k,\alpha}}  \\
             &= -\sum_{i=1}^{N_{atom}} \sum_{j=1}^{M_{atom,i}} \frac{\partial e_{MLFF,i}}{\partial \lambda_{atom,i,j}} \frac{\partial \lambda_{atom,i,j}}{\partial R_{k,\alpha}}
        \end{aligned}
        \label{eq:MLFFforce}
        \end{equation}
    \end{minipage} &
    \begin{minipage}{0.45\textwidth}
        \centering
        \begin{equation}
        \begin{aligned}
             &F_{k,\alpha} = -\frac{\partial E_{CEMT,sys}}{\partial R_{k,\alpha}}  \\
             &= -\sum_{i=1}^{N_{ref}} w_i \sum_{j=i}^{M_{ref,i}} \frac{\partial e_{CEMT,i}}{\partial \lambda_{ref,i,j}} \frac{\partial \lambda_{ref,i,j}}{\partial R_{k,\alpha}}
        \end{aligned}
        \label{eq:CEMTforce}
        \end{equation}
    \end{minipage} \\
  \end{tabular}
\label{tab:EquationComparison2}
\caption{Basic equations for force calculations based on the theories of Machine Learning Force Fields (MLFF), Chemical Environment Modeling Theory (CEMT).  $E_{sys}$ denotes system level total energy, $e_i$ denotes energy contribution, $w_i$ denotes weight, $\lambda_{i}$ denotes a feature used to describe the local chemical environments. $F_{k,\alpha}, \alpha \in \lbrace x,y,z \rbrace$ is the force component acting on atom $k$, $R_{k,\alpha}$ is the atom position and finally $N, M$ denotes the number of reference points, and the number of features per reference point, respectively.  }
\end{table}

MLFF has gained substantial traction since their inception by Behler and Parrinello in 2012 \cite{BehlerParrinello}, primarily attributed to their speed, accuracy, and adaptability. The landscape of MLFF methods has broadened over the years, categorically falling into two main branches. On one hand, explicit featurization-based MLFF models utilize carefully designed functions to represent local chemical environments around atoms. Some examples are the ACSF \cite{BehlerParrinello}, SOAP \cite{SOAP}, Gaussian Multipole (GMP) \cite{GMP}, SNAP \cite{SNAP}, and ACE \cite{ACE}, among others. Conversely, graph-based deep learning models, such as MPNN \cite{MPNN}, MEGNet \cite{MEGNet}, CGCNN \cite{CGCNN}, CHGNet \cite{CHGNet}, NequIP \cite{NequIP}, and many more, rely on data-driven learning of atomic features, often leveraging advanced model architectures like message passing and self-attention.

At the core of the myriad MLFF approaches lies a foundational assumption: the system's total energy ($E_{MLFF,sys}$) can be conceptualized as the cumulative sum of energies localized on individual atoms ($e_{MLFF,i}$), as shown in Equation \ref{eq:MLFFPartition}, which is intuitive. Note that $i$ runs through the atoms in a system. The atomic energy contribution, $e_{MLFF,i}$, depends on the local atomic environment of the atom, represented as a set of features ($\vec{\lambda}_{atom,i}$) that are defined by the spatial arrangement of neighboring atoms $\vec{r}$ and their atom types $\vec{A}$, as shown in Equation \ref{eq:MLFFEnergyDensity}. Here, $f$ is usually approximated by a machine learning model like neural networks or Gaussian processes. Once $f$ is trained, the force components $F_{k,\alpha}, \alpha \in \lbrace x,y,z \rbrace$ acting on atom $k$, with respect to its coordinates $R_{k,\alpha}$ can be written as in Equation \ref{eq:MLFFforce}. Notice that $j$ runs through the features of each atom $i$ in the system, and $M_{atom,i}$ is the number of features for atom $i$. This formulation employs the chain rule, differentiating model $f$, and descriptors or deep learning models accordingly.

The atom-centered description, while logical and intuitive, is not the sole viable approach. Take, for instance, real-space finite difference DFT (FD-DFT) and how it calculates electronic energy. When one disregards the self-consistent field (SCF) cycle, it’s observed that the energy contributions arise from each finite difference grid point. Given the fine spatial resolution, DFT can employ relatively straightforward descriptions for each grid point, such as density and gradient of density, and the associated models (functionals) are typically straightforward, harboring fewer fitting parameters. On the other hand, traditional atomistic and coarse-grained force field (FF) models attribute energy contributions not to individual atoms but to entities like bond distance, bond angles, dihedral angles, or groups of atoms. These models, with their swiftness and acceptable accuracy, necessitate substantial fitting and parameter tuning. Recognizing these alternate approaches underscores the realization that a singular focus on atom centers as reference points could inadvertently limit the progress and advancements in MLFF models.

\section{Theory}
\label{sec:Theory}
The proposed theory - Chemical Environment Modeling Theory, or CEMT - is the generalized version of MLFF, and the generalization is made by relaxing the atom-centered description to any set of reference points in the system $\lbrace {\vec{x}_i} \rbrace$. This is shown mathematically in Equation \ref{eq:CEMTPartition}. When $\lbrace {\vec{x}_i} \rbrace$ is the set of positions of atom centers, this equation is the same as Equation \ref{eq:MLFFPartition}; when $\lbrace{\vec{x}_i}\rbrace$ is the set of finite different grid points as in FD-DFT, this equation accounts for energy contributions accordingly as in Equation~\ref{eq:DFTPartition} Similarly, when ${\vec{x}_i}$ is chosen to be one point per group of atoms, it behaves like coarse-grained FFs. The energy contribution from each reference point still depends on its local chemical environment, up to certain cutoff distance, and force components of atoms can be calculated accordingly. It is worth emphasizing how the weighting factor, $w_i$, allows for more flexibility. In traditional MLFF, it's intuitive that all atomic energy contributions are weighted equally, but that's not necessarily the case for other choices of reference points. Also, until now the features for describing the local environment of reference points are still based on the atomic environment, so any existing local atomic environment-based model can be extended to this framework. More explicitly, for featurization function-based models, one can imagine having "ghost atoms" at the reference points that are featurized. As for graph-based deep learning models, one can also consider using "virtual nodes" at the reference centers.

\begin{figure}[t]
	\centering
    \includegraphics[width=1.0\linewidth]{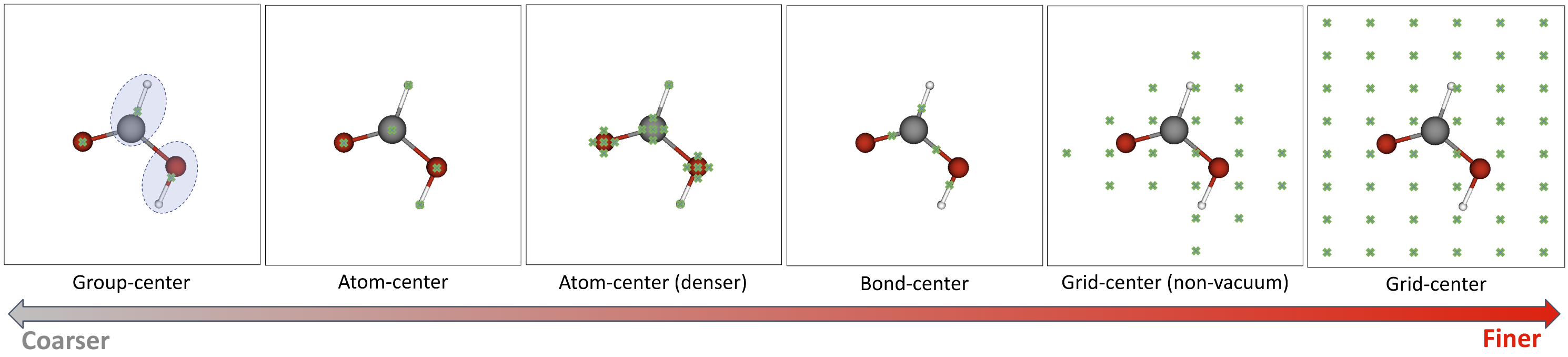}
    \caption{Some possible rational choices of reference point sets, arranged from coarser ones to finer ones.}
    \label{fig:methodSpectrum}
\end{figure}

CEMT is a simple generalization but has profound implications. First, it opens up a spectrum of new possible model architectures, as shown in Figure \ref{fig:methodSpectrum}, and different architecture would also bring different traits by design. For example, the grid-center-based models could potentially afford smaller feature sets, as well as less training data and fitting parameters. On the other hand, the group-center-based models could be faster for inference. These hypotheses are tested in subsequent experiments.

Moreover, CEMT provides a new perspective that could potentially unify the theory of MLFF with that of DFT, FF, and coarse-grained FF. Its connections with the latter two are relatively self-evident, as they are all based on describing local environments and counting energy contributions from atoms or groups. We discuss the connection with DFT in Section \ref{sec:Outlook}.

\section{Experiments}
Within the generalized CEMT framework, a series of experiments were orchestrated to test the effect of different reference points sets on models performance. To maintain uniformity, the models under evaluation in this study hinge on explicit featurization, using the Gaussian Multipole (GMP) featurization function \cite{GMP}. GMP features are fast to compute, yield fixed-length features irrespective of the elemental composition, and can be applied seamlessly to both molecular and periodic systems \cite{gmp_featurizer}.

Models are trained and benchmarked with four different reference point sets: 1) finite difference grid-centered (non-vacuum), 2) atom-centered, 3) heavy-atom-centered, and 4) bond-centered. Models and reference points are detailed further in Section \ref{sec:SI-Model}, and the model training and testing details are described in Section \ref{sec:SI-Training}. To provide comparative accuracy, we employ a uniform set of features, neural network model structures, and training procedures across all models, with the exception of the finite difference grid-center-based model which necessitated a more concise feature set and neural network model due to its extensive reference points, approximately three orders of magnitude more compared to the counterparts (Figure~\ref{fig:ref-points-qm9}). This model is also hypothesized to exhibit heightened parameter efficiency. The core aim of these experiments is not to supersede the state-of-the-art models but to demonstrate variances intrinsic to models built on distinct reference points, thereby enlightening future model development.


\begin{figure}[ht]
    \centering
    \begin{subfigure}[b]{0.7\textwidth}
        \centering
        \includegraphics[width=1.0\linewidth]{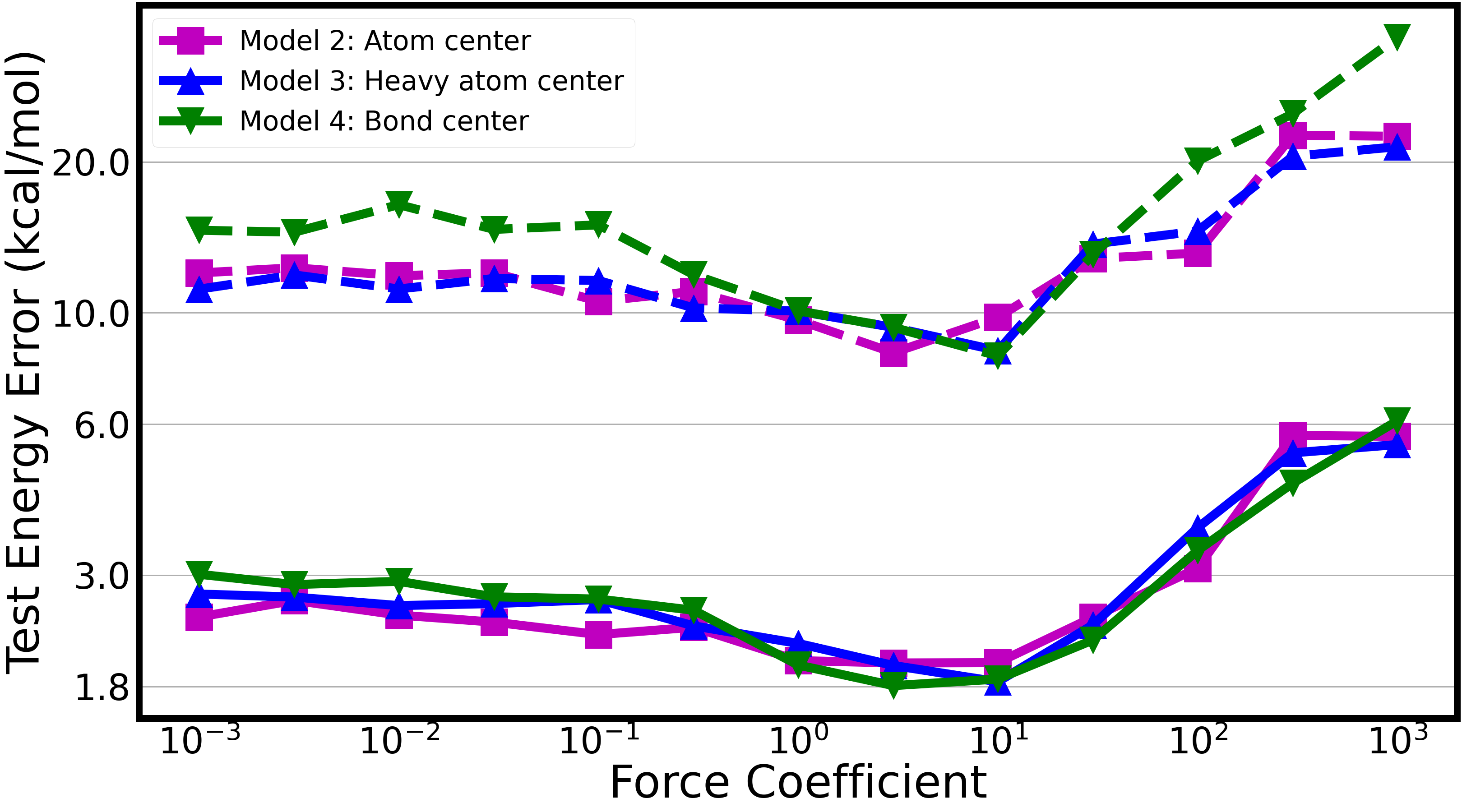}
        \caption{Energy Prediction Accuracy}
    \end{subfigure}
    \begin{subfigure}[b]{0.7\textwidth}
        \centering
        \includegraphics[width=1.0\linewidth]{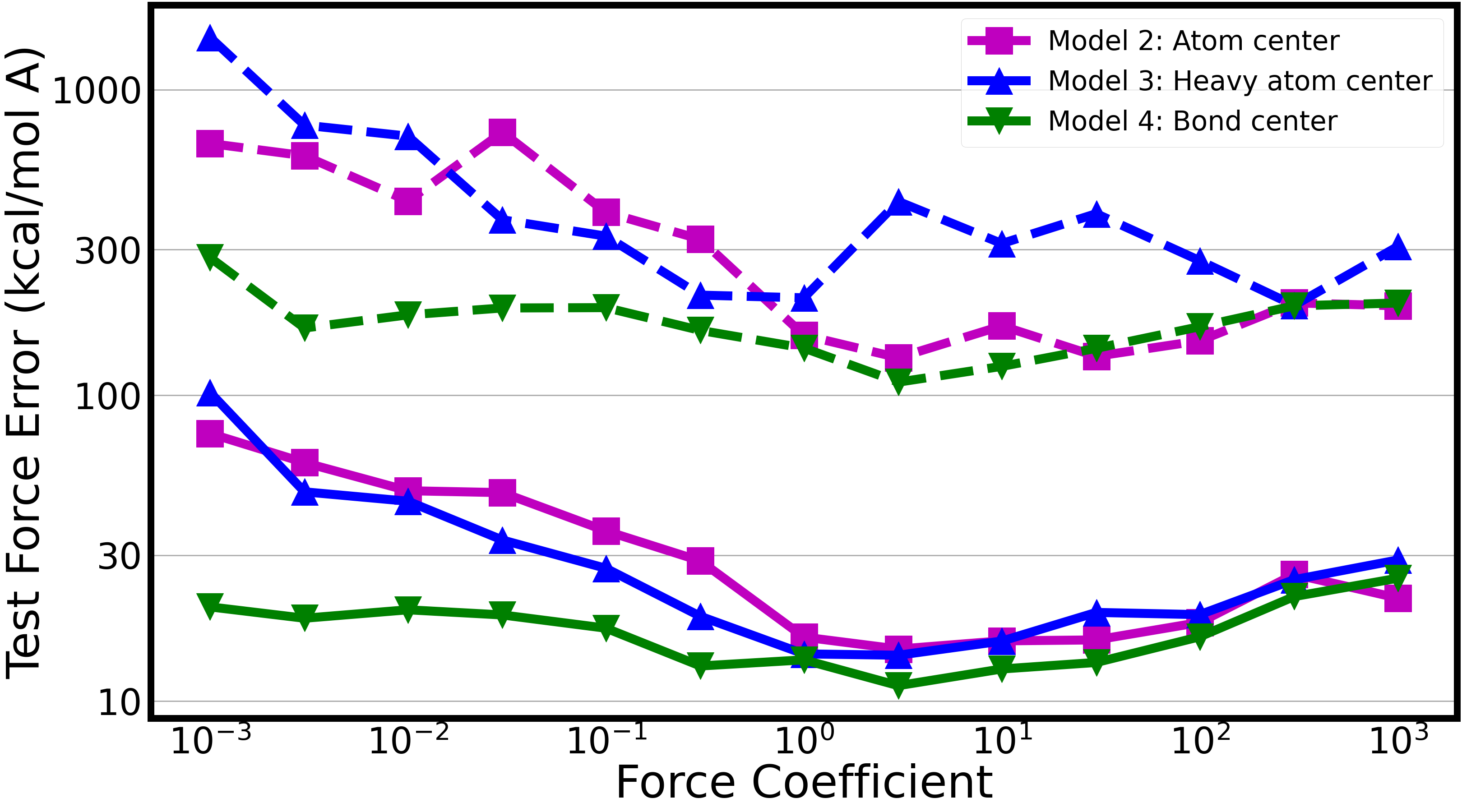}
        \caption{Energy Prediction Accuracy}
    \end{subfigure}
    \caption{Energy and force prediction results for the force-training test with different force coefficients (weight of force component during training). Solid lines are the mean absolute errors (MAEs), and dashed-lines are the max absolute errors.}
    \label{fig:Test1}
\end{figure}

\begin{figure}[ht]
    \centering
    \begin{subfigure}[t]{0.7\textwidth}
        \centering
        \includegraphics[width=1.0\linewidth]{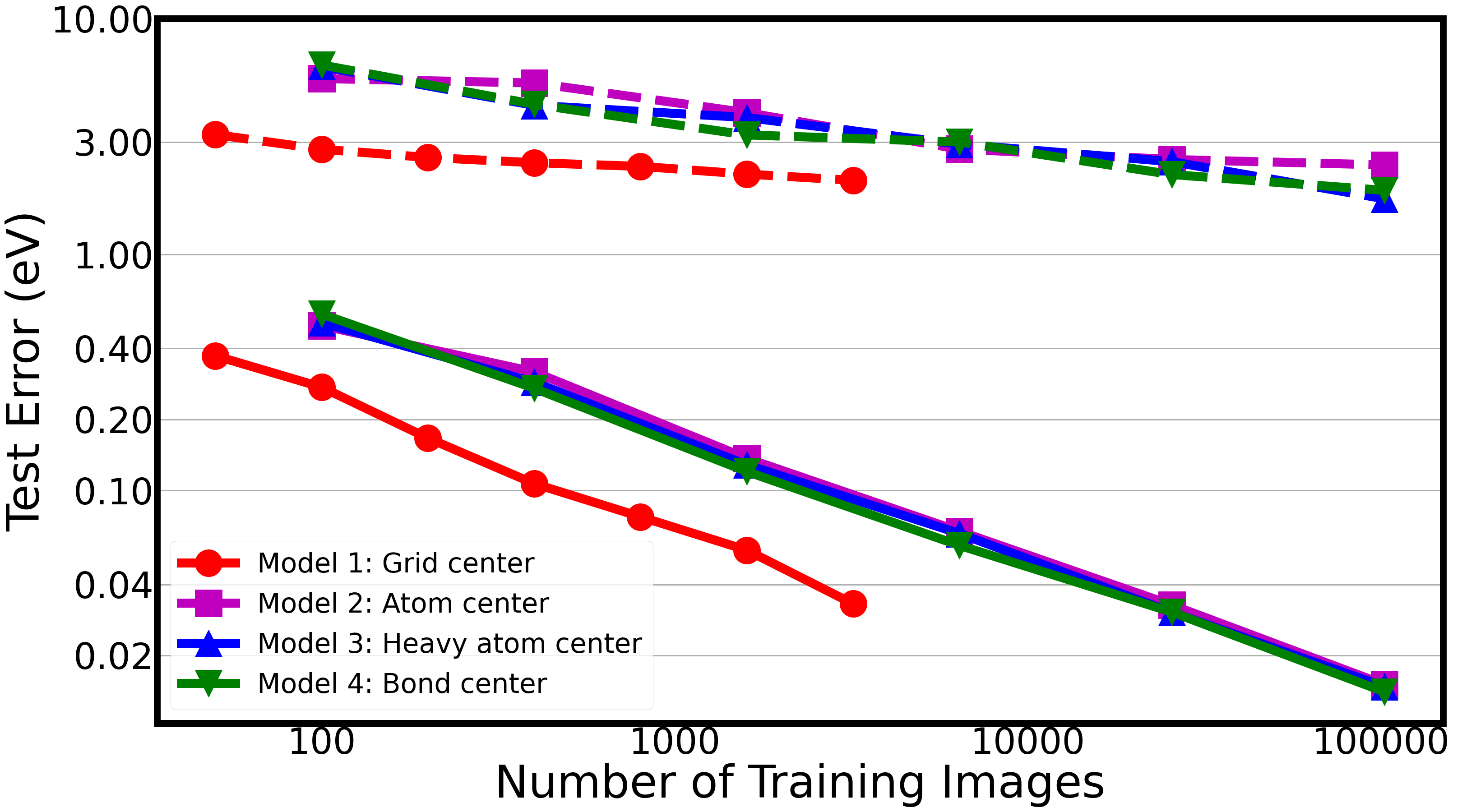}
        \caption{Learning curves of the different models. Solid lines are the mean absolute errors (MAEs), and dashed-lines are the max absolute errors.}
    \end{subfigure}
    \begin{subfigure}[t]{0.7\textwidth}
        \centering
        \includegraphics[width=1.0\linewidth]{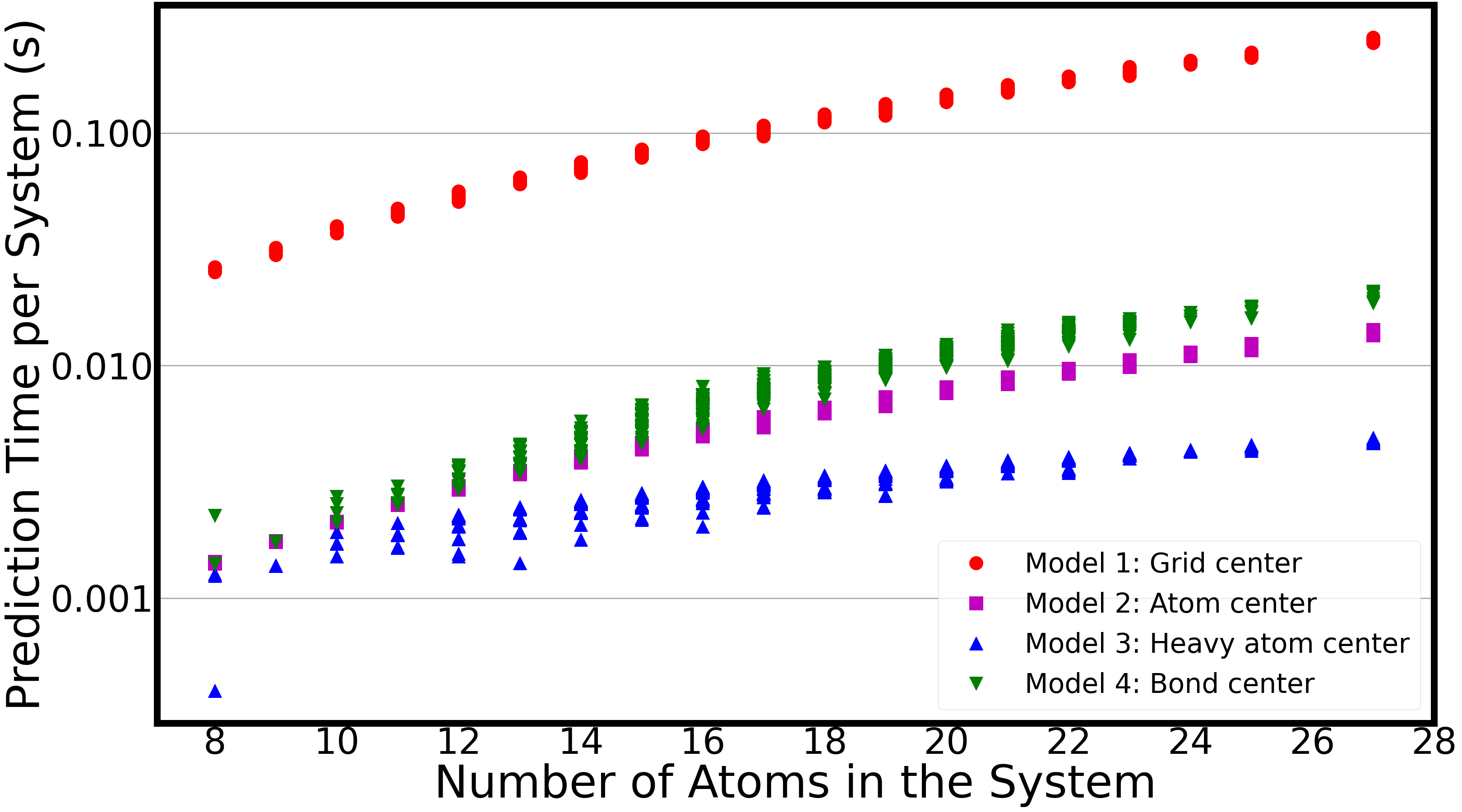}
        \caption{Prediction costs of the models, including featurization and model prediction, evaluate using 1 core.}
    \end{subfigure}
    \caption{Results for the QM9 system energy learning efficiency test}
    \label{fig:Test2}
\end{figure}

\begin{figure}[ht]
    \centering
    \begin{subfigure}[b]{0.7\textwidth}
        \centering
        \includegraphics[width=1.0\linewidth]{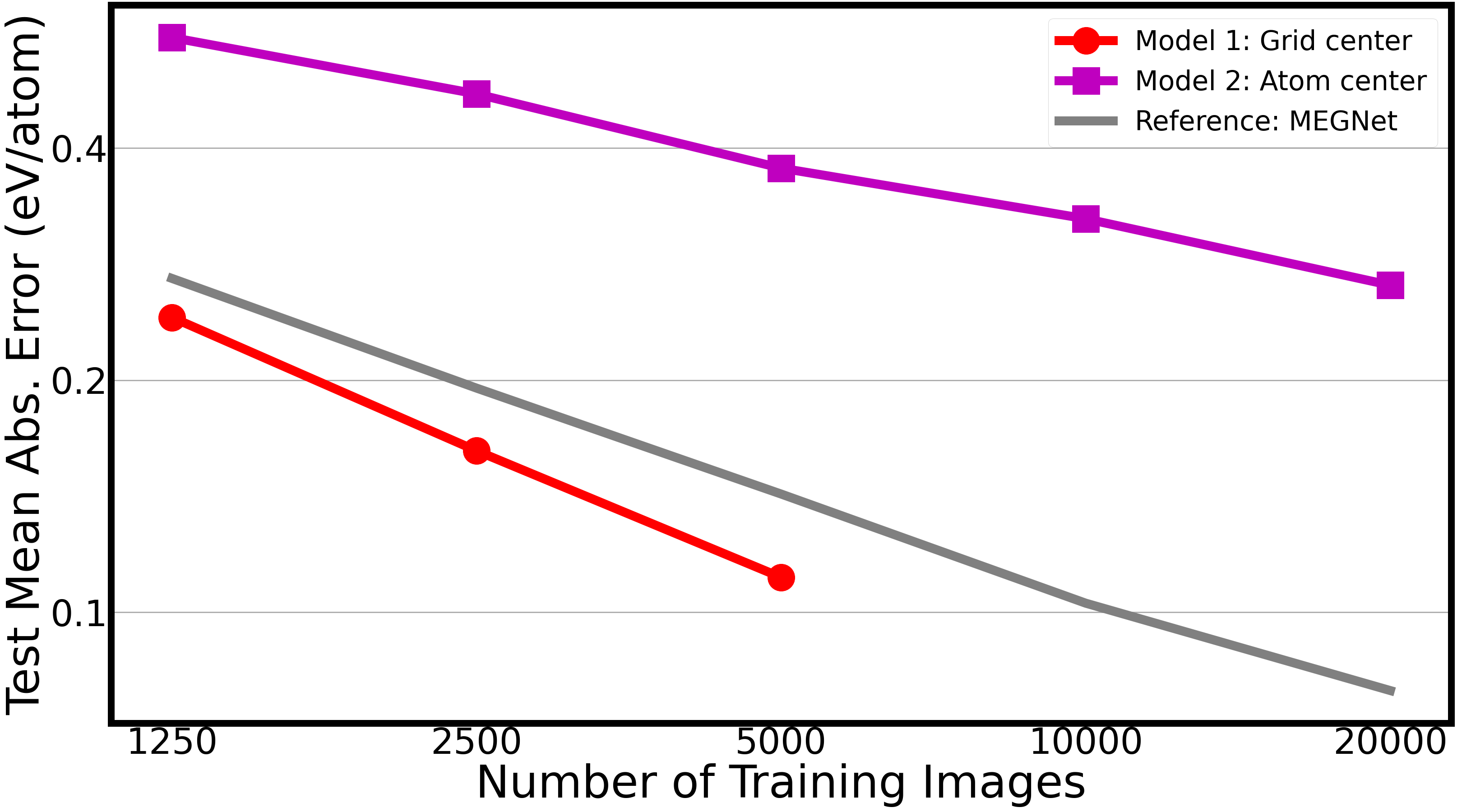}
        \caption{Learning curves for mean absolute error (MAE).}
    \end{subfigure}
    \begin{subfigure}[b]{0.7\textwidth}
        \centering
        \includegraphics[width=1.0\linewidth]{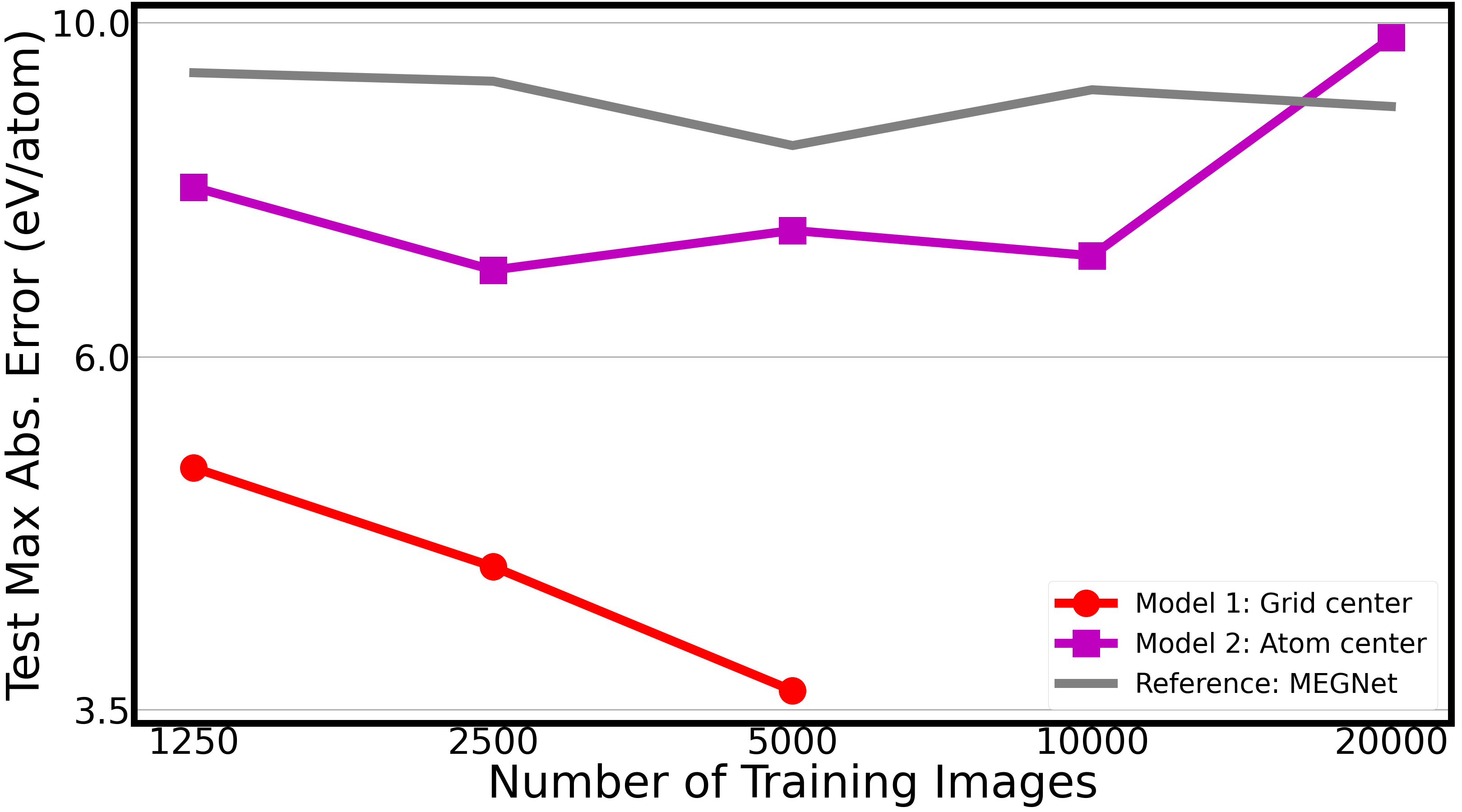}
        \caption{Learning curves for max absolute error. }
    \end{subfigure}
    \caption{Results for the MP formation energy learning efficiency test, with comparison to the graph-based MEGNet model.}
    \label{fig:Test3}
\end{figure}

The first test, a force training test, utilizes the aspirin molecular dynamics trajectory from the MD17 dataset \cite{MD17}. $1,000$ snapshots, randomly chosen, serve as the training set and are constrained by force, with the residual data deployed for testing, and results are shown in Figure \ref{fig:Test1}. The results demonstrate some variance in prediction accuracy, particularly in force prediction. But more importantly, the similarities in performance of non-atom-centered and atom-centered models corroborates the feasibility of force training with reference point sets beyond the dominant atom-centered perspective.  The grid center-based model is not included due to cost, but force-training is in theory possible.

The second test assesses the learning and inference efficiency of molecular systems, leveraging the QM9 dataset \cite{QM9}, as illustrated in Figure \ref{fig:Test2}. Remarkably, the grid-centered model, despite its reduced features per reference point and fitting parameters, shows superior data efficiency and diminished error bounds, albeit at the expense of training and inference speed. On the other hand, the heavy-atom-centered model displays comparable learning efficiency to its atom-centered counterpart but with enhanced speed owing to fewer reference points.

The final test evaluates the learning and inference efficiency of periodic systems using the Materials Project (MP) dataset \cite{MaterialsProject}, compared against the performance of the graph neural network MEGNet model \cite{MEGNet}, and with outcomes presented in Figure \ref{fig:Test3}. The heavy-atom-centered and bond-centered representations are not included since all nearly atoms are heavy atoms, and bond definition is not straightforward. The results mirror the preceding test, with the grid-centered model showing the best data efficiency. More importantly, it is robust as its error bound decreases systematically. The superiority of MEGNet compared to our atom-centered model underscores the pivotal role of model architecture, and points to room for improvement to the existing results using grid-centered reference points, through the design of deep learning model architectures.

It is evident that choice of reference points is an additional dimension to add complexity to the model, other than feature set and model architecture. However, the current model training philosophy, which aligns with fitting to the total energy of a system, escalates the training difficulty as the average number of reference points per system increases. Consequently, the grid-centered reference point model posed substantial challenges in training, necessitating extensive disk space, GPU memory, and training time, notwithstanding its data efficiency. In the next section, we discuss how a unique connection between CEMT and DFT may herald the prospect of a more effective training protocols.

\section{Outlook}
\label{sec:Outlook}

Deeper connections can be made between DFT and CEMT, other than how energy contributions are partitioned. One can start by considering real-space orbital-free finite element DFT (FE-DFT), whose formulation can be written as in Equations \ref{eq:DFTPartition} and \ref{eq:DFTEnergyDensity}. Note that the features describing local chemical environments ($\vec{\lambda}_{grid,i}$) depend on local electron density distribution ($\rho$), which in turn is a function of the relative positions and types of neighboring atoms. 

%

On the other hand, let's consider CEMT models utilizing special featurization functions that describe local chemical environments via featurizing the approximated electronic environments, such as the Gaussian Multipole (GMP) function. For these methods, each feature $\lambda$ is a function of the approximated distribution of the electron density: 
\begin{equation}
    \lambda = \lambda (\hat{\rho}(\vec{r}, \vec{A}))
\end{equation}
where $\hat{\rho}$ is the approximated electron density distribution via linear combinations of basis functions ($\phi$), which depends on $\vec{r}$ and $\vec{A}$. In the case of GMP, $\hat{\rho}$ is a linear combination of atom-centered Gaussian functions ($G_{dens,j,k}$ ).

\begin{equation}
\hat{\rho}_{GMP} = \sum_{\substack{l\\ basis}} a_l \phi_{dens,l}= \sum_{\substack{j\\atoms}} \sum_{\substack{k\\ Gaussians}} a_{j,k} G_{dens,j,k}
\label{eq:CEMTDensity}
\end{equation}

where $a$ are the coefficients of basis functions. Hence, in this special case, the energy contributions can be written as

\begin{equation}
    e^{*}_{CEMT,i} = f \circ  \vec{\lambda}_{ref,dens,i}\circ  \hat{\rho}(\vec{a};\vec{r},\vec{A})
\label{eq:CEMTEnergyDensityGMP}
\end{equation}

which is exactly the same as that of FE-DFT in Equation \ref{eq:DFTEnergyDensity}. Further, this formulation enables variational optimization of the coefficients $a$ for the optimization of electron density distribution, akin to SCF cycles in DFT calculations.

\begin{equation}
    \frac{\partial E_{CEMT,sys}}{\partial a_{j,k}}   = \sum_{i=1}^{N_{ref}} \frac{\partial e_{CEMT,i}}{\partial a_{j,k}}
    = \sum_{i=1}^{N_{ref}} \sum_{j'=i}^{M_{ref,i}} \frac{\partial e_{CEMT,i}}{\partial \lambda_{ref,i,j'}} \frac{\partial \lambda_{ref,i,j'}}{\partial a_{j,k}}
\label{eq:MLFFDFTVariation}
\end{equation}

It is worth to emphasize that, due to this connection between the CEMT-enabled model (especially when choosing grid centers as reference points) and FE-DFT, it is foreseeable that the training of these models could potentially leverage several orders of magnitudes more information from training data prepared by FE-DFT calculations. More specifically, the spatially-resolved energy densities and charge densities can be utilized, instead of just total energies. With the grid-center-based models, which are already data-efficient, it is hopeful that utilizing the grid-point resolved information would further improve their data-efficiency, potentially by orders or magnitudes, and the model training will be less challenging. Finally, it is also worth exploring that CEMT-based MLFF models could potentially further take inspirations from how DFT partitions energy contribution into kinetic, Coulombic, external, and exchange-correlation contributions. This could be a route to build known quantum-mechanical laws and limits into the architecture of the ML models, which would potentially improve their robustness and data-efficiency.

\bibliographystyle{unsrt}  
\bibliography{main}  

\begin{thebibliography}{10}

\bibitem{BehlerParrinello}
J\"org Behler and Michele Parrinello.
\newblock Generalized neural-network representation of high-dimensional potential-energy surfaces.
\newblock {\em Phys. Rev. Lett.}, 98:146401, 4 2007.

\bibitem{SOAP}
Albert~P Bartók, Sandip De, Carl Poelking, Noam Bernstein, James~R Kermode, Gábor Csányi, and Michele Ceriotti.
\newblock Machine learning unifies the modeling of materials and molecules.
\newblock {\em Science advances}, 3(12):e1701816--e1701816, 2017.

\bibitem{GMP}
Xiangyun Lei and Andrew~J. Medford.
\newblock A universal framework for featurization of atomistic systems.
\newblock {\em The Journal of Physical Chemistry Letters}, 13(34):7911--7919, 2022.
\newblock PMID: 35980312.

\bibitem{SNAP}
A.P. Thompson, L.P. Swiler, C.R. Trott, S.M. Foiles, and G.J. Tucker.
\newblock Spectral neighbor analysis method for automated generation of quantum-accurate interatomic potentials.
\newblock {\em Journal of Computational Physics}, 285:316--330, 2015.

\bibitem{ACE}
Ralf Drautz.
\newblock Atomic cluster expansion for accurate and transferable interatomic potentials.
\newblock {\em Phys. Rev. B}, 99:014104, Jan 2019.

\bibitem{MPNN}
Justin Gilmer, Samuel~S Schoenholz, Patrick~F Riley, Oriol Vinyals, and George~E Dahl.
\newblock Neural message passing for quantum chemistry.
\newblock {\em ArXiv}, 2017.

\bibitem{MEGNet}
Chi Chen, Weike Ye, Yunxing Zuo, Chen Zheng, and Shyue~Ping Ong.
\newblock Graph networks as a universal machine learning framework for molecules and crystals.
\newblock {\em Chemistry of Materials}, 31(9):3564--3572, 2019.

\bibitem{CGCNN}
Tian Xie and Jeffrey~C Grossman.
\newblock Crystal graph convolutional neural networks for an accurate and interpretable prediction of material properties.
\newblock {\em Physical review letters}, 120(14):145301--145301, 2018.

\bibitem{CHGNet}
Bowen Deng, Peichen Zhong, KyuJung Jun, Janosh Riebesell, Kevin Han, Christopher~J. Bartel, and Gerbrand Ceder.
\newblock Chgnet as a pretrained universal neural network potential for charge-informed atomistic modelling.
\newblock {\em Nature Machine Intelligence}, page 1–11, 2023.

\bibitem{NequIP}
Simon Batzner, Albert Musaelian, Lixin Sun, Mario Geiger, Jonathan~P. Mailoa, Mordechai Kornbluth, Nicola Molinari, Tess~E. Smidt, and Boris Kozinsky.
\newblock E(3)-equivariant graph neural networks for data-efficient and accurate interatomic potentials.
\newblock {\em Nature Communications}, 13(1):2453, 2022.

\bibitem{gmp_featurizer}
Xiangyun Lei and Joseph Montoya.
\newblock Gmp-featurizer: A parallelized python package for efficiently computing the gaussian multipole features of atomic systems.
\newblock {\em Journal of Open Source Software}, 8(88):5476, 2023.

\bibitem{MD17}
Stefan Chmiela, Alexandre Tkatchenko, Huziel~E. Sauceda, Igor Poltavsky, Kristof~T. Schütt, and Klaus-Robert Müller.
\newblock Machine learning of accurate energy-conserving molecular force fields.
\newblock {\em Science Advances}, 3(5):e1603015, 2017.

\bibitem{QM9}
Raghunathan Ramakrishnan, Pavlo~O Dral, Matthias Rupp, and O.~Anatole von Lilienfeld.
\newblock Quantum chemistry structures and properties of 134 kilo molecules.
\newblock {\em Scientific data}, 1(1):140022--140022, 2014.

\bibitem{MaterialsProject}
Anubhav Jain, Shyue~Ping Ong, Geoffroy Hautier, Wei Chen, William~Davidson Richards, Stephen Dacek, Shreyas Cholia, Dan Gunter, David Skinner, Gerbrand Ceder, and Kristin~A. Persson.
\newblock {Commentary: The Materials Project: A materials genome approach to accelerating materials innovation}.
\newblock {\em APL Materials}, 1(1):011002, 07 2013.

\end{thebibliography}

\newpage 

\setcounter{section}{0} 
\setcounter{equation}{0} 
\setcounter{figure}{0} 
\setcounter{table}{0} 
\setcounter{page}{1} 

\renewcommand{\thefigure}{S\arabic{figure}} 
\renewcommand{\thetable}{S\arabic{table}} 
\renewcommand{\theequation}{S\arabic{equation}} 
\renewcommand{\thesection}{S\arabic{section}} 
\renewcommand{\thepage}{S\arabic{page}} 

\section{Supplementary Information: The Role of Reference Points in Machine-Learned Atomistic Simulation Models}

\subsection{Model and reference points details}
\label{sec:SI-Model}
As mentioned, 4 models with 4 difference choices of reference point sets are tested: 
1) Finite difference grid-centered (non-vacuum) 2) Atom-centered 3) Heavy-atom-centered 4) Bond-centered

Here are some more details for how the reference point sets are defined
\begin{itemize}
    \item Model 1: reference points are defined as the grid points of an uniform finite difference. The grid has 0.3 $\angstrom$ grid spacing. Grid points that are at least 3.0 $\angstrom$ away from any atoms are discarded. Weight of the energy contribution of each grid point (as defined in Equation \ref{eq:CEMTPartition}) is $0.3 ^ 3$. Note that for the periodic systems, it is not possible to ensure the grid spacing is exactly $0.3 \angstrom$, and the cells are rarely orthogonal. Therefore, the weights in this case depend on the actual volume of the grid.
    \item Model 2: reference points are defined as the atom positions.
    \item Model 3: reference points are defined as the atom positions for atoms other than hydrogen.
    \item Model 4: reference points are defined as the central points of the bonds. If two atoms are within 2 $\angstrom$ distance from each other, it is considered there is a bond in this simple experiment.
\end{itemize}

Here are details of the feature set for different models, and the architecture of the SNNs (see~\ref{sec:SI-SNN}). Activation function is set to be the GeLU function, and batch normalization is also performed at each layer. For the detailed meaning of ``Radial Sigmas'' and ``Angular Orders'', please refer to the original publications of the Gaussian Multipole featurization function \cite{GMP}, and the computation package \cite{gmp_featurizer}.

\begin{table}[h!]
\centering
\begin{tabular}{|c|c|c|c|}
\hline
Model & Radial Sigmas & Angular Orders & \# Features  \\
\hline
1 & [0.1, 0.2, 0.3, 0.4, 0.5] & [-1,0,1,2,3] & 21 \\
2 & [0.05, 0.1, 0.15, ..., 1.95, 2.0] & [-1,0,1,2,3,4,5] & 241  \\
3 & [0.05, 0.1, 0.15, ..., 1.95, 2.0] & [-1,0,1,2,3,4,5] & 241  \\
4 & [0.05, 0.1, 0.15, ..., 1.95, 2.0] & [-1,0,1,2,3,4,5] & 241 \\
\hline
\end{tabular}
\caption{Details of the feature sets of the tested models}
\label{tab:feature_detail}
\end{table}

\begin{table}[h!]
\centering
\begin{tabular}{|c|c|c|}
\hline
Model & Model Hidden Layer & \# Parameters \\
\hline
1 & (128,128,64,64,32,32,16,16) & 36,657 \\
2 & (512,256,128,64,32) & 300,481 \\
3 & (512,256,128,64,32) & 300,481 \\
4 & (512,256,128,64,32) & 300,481 \\
\hline
\end{tabular}
\caption{Details of the tested models}
\label{tab:model_detail}
\end{table}

\begin{figure}[ht]
    \centering
    \begin{subfigure}[b]{0.48\textwidth}
        \centering
        \includegraphics[width=1.0\linewidth]{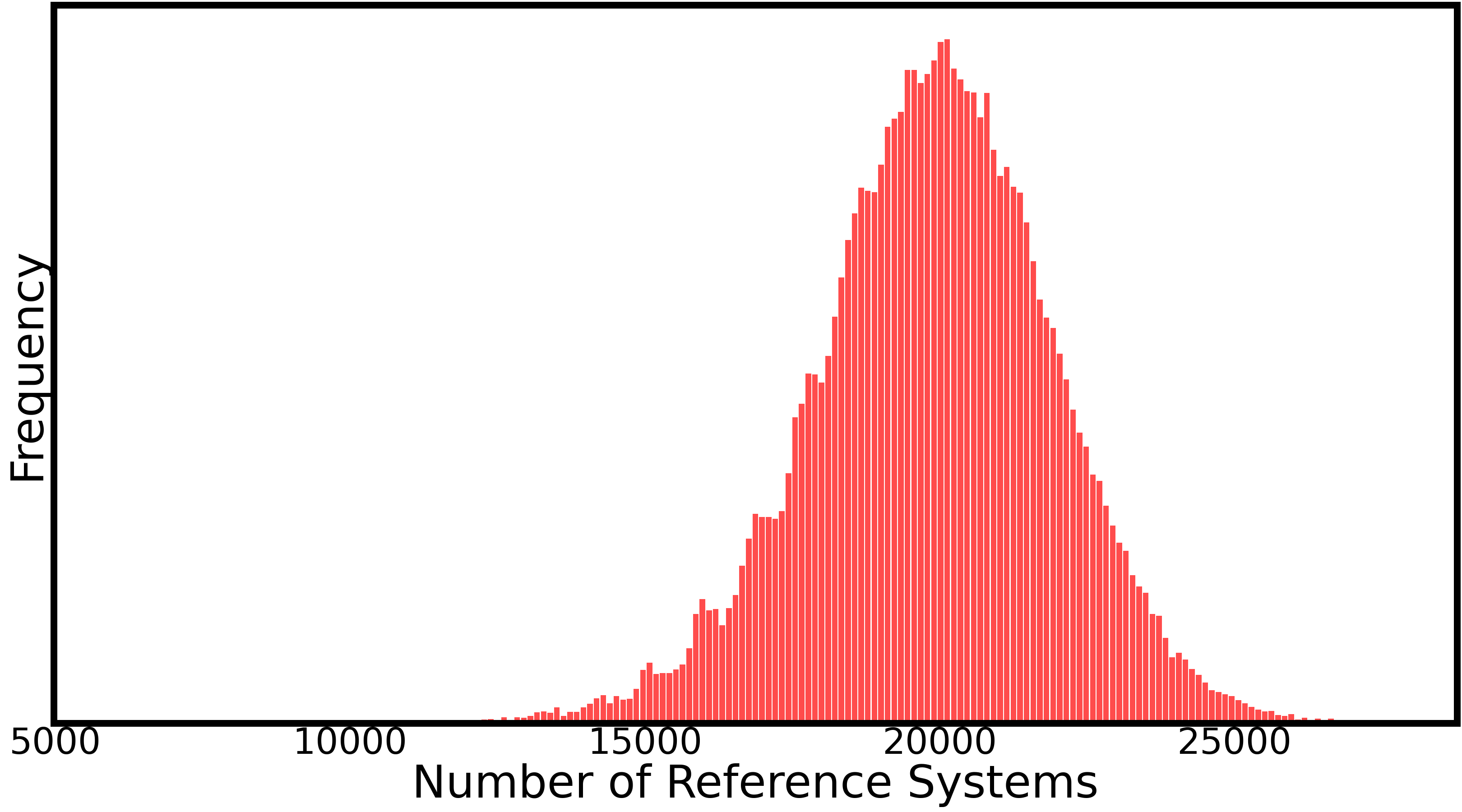}
        \caption{Finite difference grid-centered}
    \end{subfigure}
    \hfill
    \begin{subfigure}[b]{0.48\textwidth}
        \centering
        \includegraphics[width=1.0\linewidth]{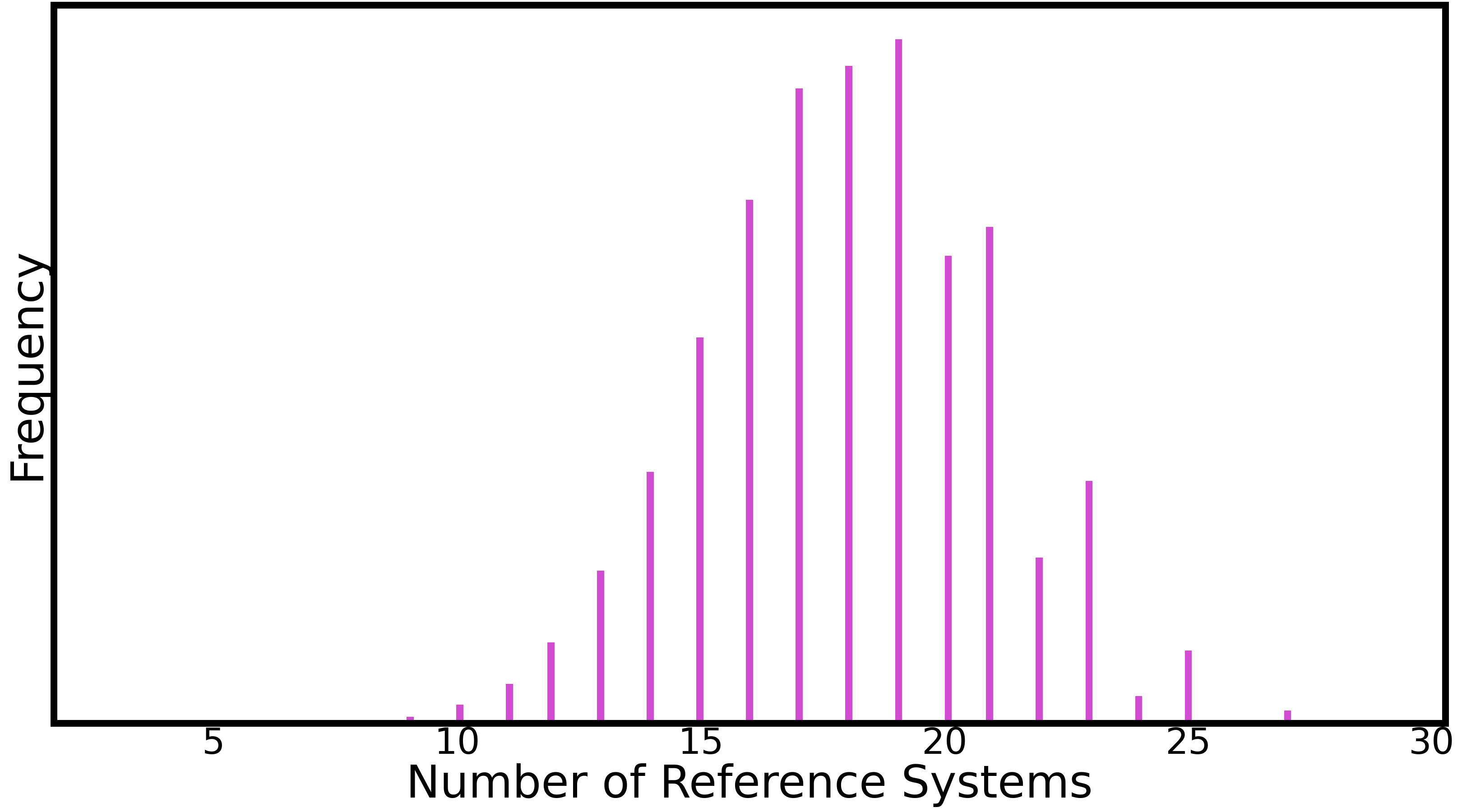}
        \caption{Atom-centered}
    \end{subfigure}
    \caption{Results for the MP formation energy learning efficiency test, with comparison to the graph-based MEGNet model.}
    \begin{subfigure}[b]{0.48\textwidth}
        \centering
        \includegraphics[width=1.0\linewidth]{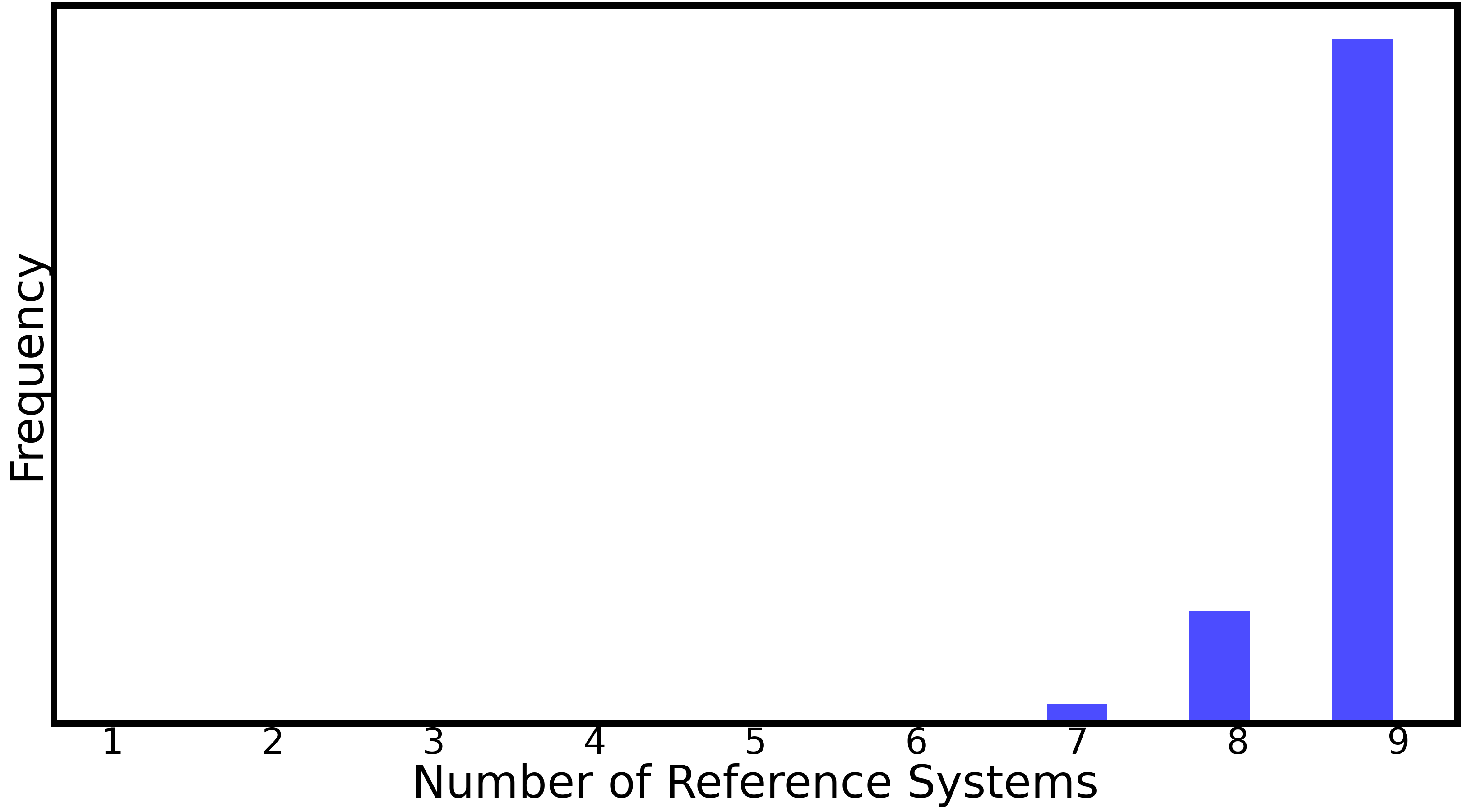}
        \caption{Heavy-atom-centered}
    \end{subfigure}
    \hfill
    \begin{subfigure}[b]{0.48\textwidth}
        \centering
        \includegraphics[width=1.0\linewidth]{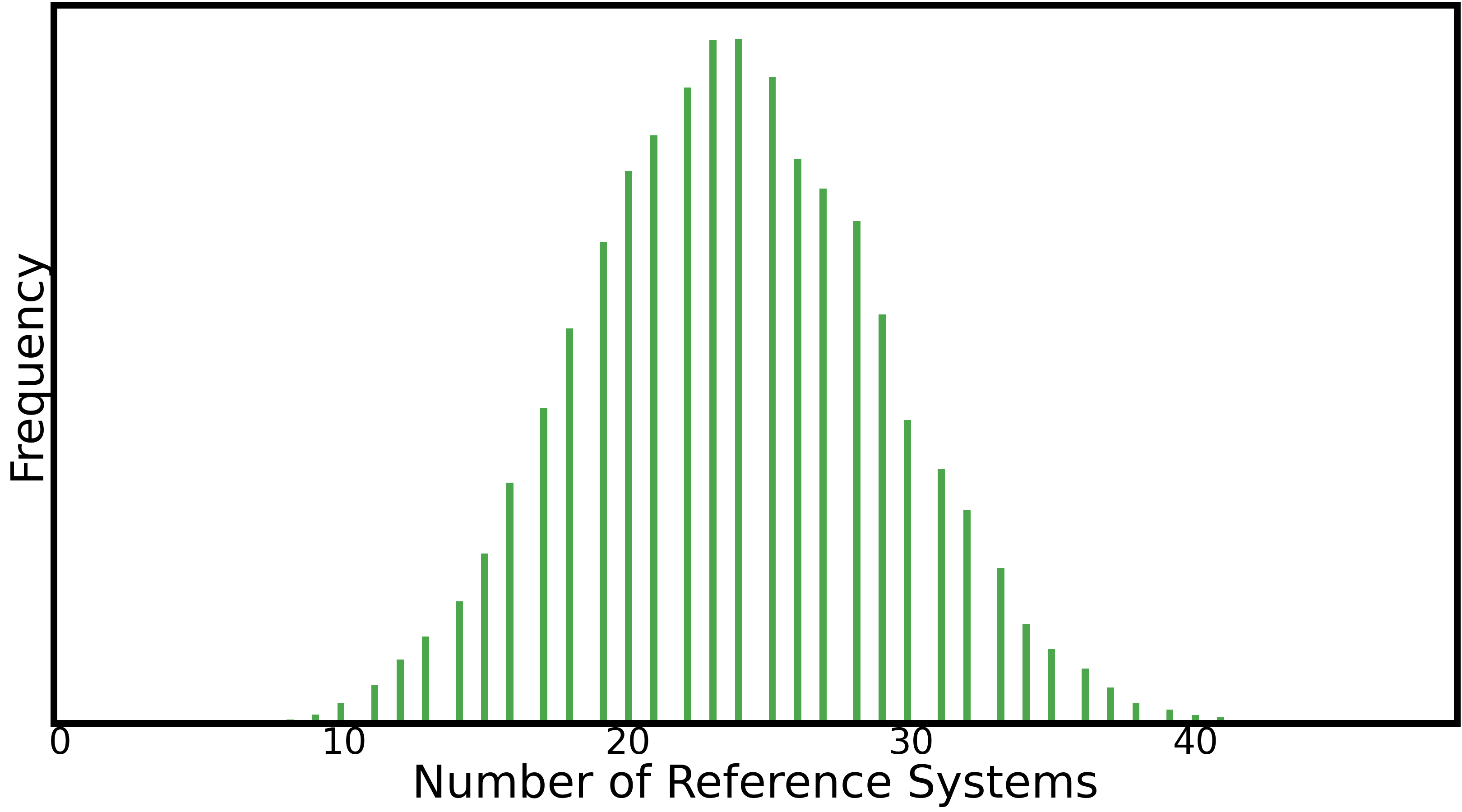}
        \caption{Bond-centered}
    \end{subfigure}
    \caption{Count of reference points for each molecule's representation in the QM9 training task.}
    \label{fig:ref-points-qm9}
\end{figure}

The numbers of reference points for the 4 choices are counted and plotted in Figure~\ref{fig:ref-points-qm9}. The dataset is the QM9 dataset.

\subsection{Single Neural Network Architecture}
\label{sec:SI-SNN}
Single Neural Network (SNN) architecture serves as a distinct approach that deviates from the conventional High-Dimensional Neural Network (HDNN) architecture. HDNN is generally utilize by NN-based MLFF models that uses atom-distinct featurization functions like ACSF and SOAP, and it has distinct neural networks for different atom types  to predict energy contributions from atomic features. The SNN model, on the other hand, uses a single shared neural network for all atom types. In the context of CEMT, the SNN are shared across all reference points for energy contribution prediction.

\subsection{Training and Test Details}
\label{sec:SI-Training}
\subsubsection{Data Preparation}
To prepare the training data for all tests, they were first downloaded from their corresponding websites, and all converted to ASE Atoms objects.

The corresponding training datasets was formed by randomly selected the needed amount of systems for training, and the rest is used for testing.

For the QM9 and MP datasets, it was checked that all the training sets contain all possible elements, and a linear model was fitted to each training set to calculate the per-element biases for each elements (details in the original GMP publication \cite{GMP})

The GMP features for each dataset are computed as listed in the previous section. Note that data augmentation is applied to the grid center feature models because the definition of the grid is not unique. It was augmented by starting with a reference grid, and move it 0.15 \angstrom in x, y and z axis combinatorially for another 7 grids.

\subsubsection{MD17 Model Training Procedure}
For all three tested models (Models 2, 3, 4), the training procedure was the same. The SNN model was initiated with the Xavier initialization scheme, and then trained for 500 epochs. The learning rate started at $1^-3$, and was decreased by a factor of 2 after every 100 epochs. The batchsize is 64. Note that, since it's force training, the numerical precision of the models was set to be double precision.

\subsubsection{QM9 Model Training Procedure}
For Model 1, the SNN model was initiated with the Xavier initialization scheme, and then trained for 1000 epochs. The learning rate started at $1^{-3}$, and was decreased by a factor of 2 after every 200 epochs. However, some manually adjustments were also needed. The batchsize was 512, and 10 \% data was used for validation during training.

For the other three tested models (Models 2, 3, 4), the training procedure was the same. The SNN model was initiated with the Xavier initialization scheme, and then trained for 2000 epochs. The learning rate started at $1^{-3}$, and was decreased by a factor of 2 after every 400 epochs. The batchsize was 256, and 10 \% data was used for validation during training.

The numerical precision of all models was set to be single precision.

\subsubsection{MP Model Training Procedure}

For Model 1, the SNN model was initiated with the Xavier initialization scheme, and then trained for 1000 epochs. The learning rate start at $1^{-3}$, and was decreased by a factor of 2 after every 200 epochs. However, some manually adjustments were also needed. The batchsize was 1024, and 10 \% data was used for validation during training.

For the other three tested models (Models 2, 3, 4), the training procedure was the same. The SNN model was initiated with the Xavier initialization scheme, and then trained for 2000 epochs. The learning rate started at $1^{-3}$, and was decreased by a factor of 2 after every 400 epochs. The batchsize was 256, and 10 \% data was used for validation during training.

The numerical precision of all models was set to be single precision.

\subsubsection{Test Procedure}
For all model tests, the models are tested against the rest of the corresponding dataset, and the results are recorded. In other words, the recorded errors are the out-of-sample test errors.

\end{document}